\documentstyle[aps,prb,epsfig,twocolumn]{revtex}

\newcommand{\bfg}[1] {\bbox{#1}}

\begin{document}

\wideabs{


\title{Optical creation of vibrational intrinsic localized modes in
anharmonic lattices\\
with realistic interatomic potentials}

\author{T. R\"{o}ssler}
\address{Infineon Technologies AG, MP CAD,
P.O. Box 800949, D-81609 Munich, Germany}

\author{J. B. Page}
\address{Department of Physics and Astronomy, Arizona State
University, Tempe, Arizona 85287-1504}


\maketitle

\begin{abstract}
Using an efficient optimal control scheme to determine the exciting
fields, we theoretically demonstrate the optical creation of 
vibrational intrinsic localized modes (ILMs) in anharmonic perfect
lattices with realistic interatomic potentials. For systems with 
finite size, we show that ILMs can be excited directly by applying a 
sequence of femtosecond visible laser pulses at THz repetition rates.
For periodic lattices, ILMs can be created indirectly via decay of an 
unstable extended lattice mode which is excited optically either by 
a sequence of pulses as described above or by a single picosecond 
far-infrared laser pulse with linearly chirped frequency. In light 
of recent advances in experimental laser pulse shaping capabilities, 
the approach is experimentally promising.
 
\end{abstract}

}
 
\section{Introduction}

Over the past several years, theoretical studies of the dynamics of 
anharmonic periodic lattices have established the existence of
intriguing vibrational excitations, characterized by well-localized
displacement patterns.\cite{ILM} These so-called intrinsic localized 
modes (ILMs) can exist at any site in a perfect lattice, in contrast to 
localized impurity modes in harmonic defect crystals. While vibrational 
ILMs with atomic scale localization have been obtained for increasingly 
realistic lattice-dynamical models,\cite{Bonart95,Kiselev96} their 
experimental study has been impeded by the lack of direct methods for their
excitation and verification. Recently, ILMs in a complex quasi-1D 
charge-density wave system were inferred from resonance Raman data, through
the use of a coupled electron-vibration model restricted to a single 
repeat unit.\cite{Swanson99}

Previously we have demonstrated theoretically that driven
ILMs can exist as a steady-state response to an applied spatially
homogeneous monochromatic driving force.\cite{Roessler95} As a natural 
extension of that work, and to address a key 
experimental question, we here describe theoretically an avenue for
the transient optical creation of ILMs: we show how they can be produced
in a 1D model lattice with realistic potentials by means of
laser pulses whose time dependence is designed by an efficient optimal
control scheme. 

Owing to their high power densities, lasers are attractive sources for
exciting large-amplitude ILMs. Indeed, there are some notable examples
of experiments with powerful lasers in the regime of anharmonic lattice
vibrations. For instance, phonon resonances measured in the ferroelectric
LiNbO$_3$ by experiments using single visible laser pulses with a duration
of 60 fs and an energy per pulse of 5 $\mu$J were interpreted in terms of
``overtones'' of the very anharmonic lowest-energy TO phonon of $A_1$
symmetry.\cite{Bakker94} Also, experiments on Ti$_2$O$_3$ using single
visible pulses with 10 nJ energy and 70 fs widths to excite the $A_{1g}$
mode apparently resulted in atomic displacements of about 0.07 \AA\ --
corresponding to 2\% of the interatomic spacing -- and revealed anharmonic
behavior.\cite{Cheng93}

Of more direct relevance here is the fact that experimental laser pulse
shaping techniques \cite{Kawashima95,Knippels95} provide considerable
flexibility in the time dependence of the applied force. Indeed, the use of
tailored fields for vibrational excitation has attracted much
interest, mainly in the context of optical control of dissociation and
reactions in molecular chemistry,\cite{Chem} but also
for the selective excitation of optical phonons in time-domain 
spectroscopy.\cite{Condmat,Weiner91} We will focus on two methods for
the transient optical creation of ILMs: (i) impulsive stimulated Raman
scattering (ISRS) excitation \cite{Dhar94} by a sequence of femtosecond
pulses at THz repetition rates from a laser operating at visible or
near-visible frequencies, and (ii) infrared (IR) excitation by a picosecond
far-IR laser pulse. For both mechanisms, the system's dynamical response
can be described classically, provided the underlying laser frequency for
the ISRS case is well off resonance with vibrational and electronic
transitions. 

Since ILMs are complex, large-amplitude vibrational excitations, the external
fields necessary for their creation are likely to have a complicated time
dependence and large magnitudes. It is therefore advantageous
to determine the optimal external fields by a systematic scheme.
In engineering, the analogous task of designing the time dependence of an
external force to steer a dynamical system towards a desired target state
is a fundamental problem. Optimal control theory \cite{Lueneberger79}
provides a solution with a rigorous mathematical foundation, based on the
variational minimization of a positive objective functional. In the realm
of atomic dynamics, this approach has been successful in the
design of external electric fields for selective bond excitation in models
of small harmonic lattices \cite{Shi88,Shi90,Schwieters90} and small
anharmonic molecules.\cite{Botina95} We apply a similar scheme to the
creation of ILMs. 

The following section discusses our theoretical framework by providing the
necessary details of the optimal control scheme and describing the specific
anharmonic model lattice we use. Our numerical results concerning
the optical creation of ILMs are given in Sec.\ \ref{sec:opticalcreation},
which is divided into two parts according to the two different optical methods 
we consider. In Sec.\ \ref{sec:discussion} we address aspects of the
experimental feasibility and discuss our results. 
Section \ref{sec:conclusion} concludes the paper, and an appendix provides 
additional qualitative insight. Some of the results presented here were
summarized in a letter.\cite{Roessler97}

\section{Theoretical background}
\label{sec:basis}

Although some notable exceptions have 
appeared,\cite{Bonart95,Kiselev96} most studies of 
ILMs have considered 1D or
2D model lattices. This restriction to simpler model systems has
facilitated progress towards a theoretical understanding of
basic ILM properties, without the additional numerical
complications encountered with more realistic 3D models of 
crystals.\cite{Kiselev96} In addition to the reasonable 
expectation that many of
these basic properties will transfer to the 3D case, there is
evidence that 1D models may apply directly to some types of motion in
3D crystals. For instance, in Ref.\ \onlinecite{Bonart95} a realistic 3D
lattice-dynamical model was considered and ILMs were obtained with
displacement patterns localized along the edge of the crystal, which can be
regarded as a natural generalization of a 1D lattice. Furthermore, it is
well known that along some high symmetry directions in 3D crystals, the
harmonic lattice dynamics map onto an effective 1D model involving the
collective motion of lattice planes. We have performed preliminary studies
showing that this mapping also occurs in the anharmonic case, for certain
polarization directions.\cite{Roessler98} This point is addressed in more
detail later. Hence, for a demonstration of the optical creation of ILMs,
we consider a diatomic 1D model lattice that incorporates realistic
features, such as standard interparticle potentials and measured harmonic
properties of real crystals. 

\subsection{The system: Hamiltonian and dynamics}
\label{sec:basissystem}

For longitudinal motion in an externally driven 1D system, the
Hamiltonian is
\begin{equation}
\label{hamiltonian}
H = \sum_n \left[\frac{p_n^2}{2m_n}+
\sum_{l>0} V_{n,n-l}\left(r_n-r_{n-l}\right)- f_n^{\rm ext}(t)r_n\right],
\end{equation}
where particle $n$ has mass $m_n$, position $r_n$ and momentum $p_n$ and
interacts with particle $n-l$ via a potential $V_{n,n-l}(r)$, to be
specified below. The external force is given by $f_n^{\rm ext}(t) =
{1 \over 2}{\cal P}_n{\cal E}^2(t)$ and $f_n^{\rm ext}(t) = q_n{\cal E}(t)$
for ISRS and IR excitation, respectively. Here ${\cal E}(t)$ is the
longitudinally polarized electric field, ${\cal P}_n \equiv
(\partial {\cal P}/\partial r_n)_0$ is the electronic polarizability
derivative evaluated at the equilibrium configuration, and $q_n$ is the
effective charge. With ${\cal P}_n=(-1)^n{\cal P}$ and $q_n=(-1)^nq$, the
external forces for both excitation methods have equal magnitudes and
alternating signs:
\begin{equation}
\label{externalforce}
f_n^{\rm ext}(t)=(-1)^n{\cal F}(t).
\end{equation}
The vibrational dynamics of this 1D system are described by Hamilton's
equations
\begin{mathletters}
\label{hamiltoneqs}
\begin{equation}
\label{hamiltonr}
\dot{r}_n = {\partial H \over \partial p_n} = {p_n \over m_n}
\end{equation}
\begin{equation}
\label{hamiltonp}
\dot{p}_n = - {\partial H \over \partial r_n} = f_n(\{r_m\}) 
+ (-1)^n{\cal F}(t),
\end{equation}
\end{mathletters}
\noindent
where
\begin{eqnarray}
\label{internalforce}
f_n(\{r_m\}) \equiv - \sum_{l>0} &&\left[V'_{n,n-l}(r_n - r_{n-l})-\right.
\nonumber\\
&&\left. V'_{n+l,n}(r_{n+l} - r_{n})\right]
\end{eqnarray}
is the total internal force on particle $n$, with
$V'_{n,m}(r_n-r_m)\equiv(dV_{n,m}/dr)|_{r=r_n-r_m}$. From Eqs.\
(\ref{hamiltoneqs}) we readily obtain the equations of motion
\begin{equation}
\label{eom}
m_n\ddot{r}_n = f_n(\{r_m\}) + (-1)^n{\cal F}(t),
\end{equation}
which are more convenient for numerical simulations of the system dynamics.

\subsection{The scheme: optimal control theory}
\label{sec:basisscheme}

In the following, we describe the most important aspects of the optimal
control scheme used in this work. More details can be found in Refs.\
\onlinecite{Shi88} and \onlinecite{Shi90}, whose compact notation 
we adopt with some modifications. 
We consider a system of $N$ particles. Unless explicitly stated
otherwise, $N$-dimensional vectors are denoted by lower case bold Roman
letters and $N \times N$-dimensional matrices by upper case bold Roman
letters. Moreover in phase space, $2N$-dimensional vectors are denoted by
lower case bold Greek letters and $2N$$\times$$2N$-dimensional matrices
by upper case bold Greek letters.


We combine the $N$-dimensional vectors ${\bf r}$ and ${\bf p}$ for the
particle positions and momenta, respectively, to form a $2N$-dimensional
phase space vector 
\begin{equation}
\bfg{\xi}^T \equiv ({\bf r}^T,{\bf p}^T)
\equiv (r_1,\dots,r_N,p_1,\dots,p_N),
\end{equation}
where the superscript $T$ denotes the transpose. Equations 
(\ref{hamiltoneqs}) are then rewritten as
\begin{equation}
\label{dynamicaleq}
\bfg{\dot{\xi}} = \bfg{\phi}[\bfg{\xi}, {\cal F}(t)],
\end{equation}
with
\begin{equation}
\label{fdefinition}
\bfg{\phi}^T[\bfg{\xi}, {\cal F}(t)] \equiv
\left[({\bf M}^{-1}{\bf p})^T, {\bf f}^T +
{\bf {\tilde q}}^T{\cal F}(t)\right],
\end{equation}
where ${\bf M}$ is the $N$$\times$$N$-dimensional diagonal mass matrix
with elements $({\bf M})_{nl} \equiv m_n\delta_{nl}$, $({\bf f})_n
\equiv f_n(\{r_m\})$, and the definition of the coupling vector
${\bf {\tilde q}}$ with components $({\bf {\tilde q}})_n \equiv (-1)^n$
allows a compact description of the external force terms. With this
notation, we can rewrite Eq.\ (\ref{eom}) as
\begin{equation}
\label{eomcompact}
{\bf M}\cdot{\bf{\ddot r}}={\bf f}+{\bf {\tilde q}}{\cal F}(t). 
\end{equation}
We furthermore specify initial conditions
\begin{mathletters}
\label{eombc}
\begin{equation}
{\bf r}(t=0) = {\bf r}_i,
\end{equation}
\begin{equation}
{\bf p}(t=0) = {\bf p}_i
\end{equation}
\end{mathletters}
at $t=0$.

In order to apply optimal control methods to our problem, we first need
to define a positive functional that reflects the physical objectives
to be reached. Starting from the lattice in some initial configuration
$({\bf r}_i,{\bf p}_i)$, we want to excite a given anharmonic mode
at a specified final time $t_f$, while keeping the magnitude of the
external force within reasonable limits. Combining the target mode
positions ${\bf r}_f$ and momenta ${\bf p}_f$ into a final phase space
vector $\bfg{\xi}_f$, we define the objective functional
\begin{eqnarray}
\label{objfunc}
J[\bfg{\xi},{\cal F}(t)] = && {1 \over 2} \left[\bfg{\xi}(t_f) -
\bfg{\xi}_f\right]^T \cdot \bfg{\Psi} \cdot \left[\bfg{\xi}(t_f) -
\bfg{\xi}_f\right] + \nonumber \\
&& {1 \over 2} \psi_{\cal F}
\int_0^{t_f} dt\, {\cal F}^2(t),
\end{eqnarray}
where the nonzero elements $(\bf \Psi )_{\alpha\alpha}
\equiv \psi_{\alpha}$ of the $2N$$\times$$2N$-dimensional diagonal 
matrix $\bf \Psi$ are positive weight factors, as is $\psi_{\cal F}$.
Note that in order for all terms in the objective functional to have
the same units, not all of the weight factors can be unitless.
Furthermore, only trajectories $\bfg{\xi}(t)$ satisfying the equations 
of motion (\ref{dynamicaleq}) are admissible during the optimization.
Including this as a constraint in the objective functional, we obtain 
the modified objective functional
\begin{equation}
\label{modobjfunc}
\bar{J}[\bfg{\xi},{\cal F}(t)] = J[\bfg{\xi},{\cal F}(t)] -
\int_0^{t_f} dt\, \bfg{\lambda}^T\cdot \left\{ {\bfg{\dot{\xi}} - 
\bfg{\phi}[\bfg{\xi}, {\cal F}(t)]} \right\},
\end{equation}
where $\bfg{\lambda}^T = (\lambda_1,\dots,\lambda_{2N})$
is a $2N$-dimensional vector of time-dependent Lagrange multipliers. 
This modified objective functional is minimized with respect to
the trajectories ${\bfg \xi}$ and external force $F(t)$ to 
obtain the optimal force. For clarity, these quantities are treated
separately in the following two paragraphs.

Variational minimization with respect to the trajectories 
$\{\xi_{\alpha}(t)\}$, including an integration by parts of the 
term $\int_0^{t_f} dt\, {\bfg \lambda}^T \cdot 
\delta \bfg{\dot{\xi}}\,$, yields dynamical equations for the 
Lagrange multipliers
\begin{equation}
\label{dynamicallagrange}
\bfg{\dot{\lambda}} + {\bf \Phi} \cdot \bfg{\lambda} = 0,
\end{equation}
where $\bf \Phi$ is a $2N$$\times$$2N$-dimensional
time-dependent matrix with elements $({\bf \Phi})
_{\alpha\beta} \equiv \partial \phi_{\beta}[\bfg{\xi},
{\cal F}(t)] / \partial\xi_{\alpha}$.
At the final time $t_f$, these dynamical equations are subject to
boundary conditions
\begin{equation}
\label{finallagrange}
\bfg{\lambda}(t_f) = \bfg{\Psi} \cdot [\bfg{\xi}(t_f) - \bfg{\xi}_f].
\end{equation}
For our definition of ${\bfg \phi}[\bfg{\xi},
{\cal F}(t)]$ [see Eq.\ (\ref{fdefinition})], the matrix
${\bf \Phi}$ decomposes into four $N$$\times$$N$-dimensional
blocks:
\begin{equation}
{\bf \Phi} =
\left(
\begin{array}{c|c}
{\bf 0} & -{\bf K} \\ \hline
{\bf M}^{-1} & {\bf 0}
\end{array}
\right),
\end{equation}
where we have defined the $N$$\times$$N$-dimensional time-dependent,
symmetric dynamical matrix ${\bf K}$ with elements
$\left({\bf K}\right)_{nl} \equiv - \partial f_n(\{r_m\})/\partial r_l$.
We can now simplify the description by considering separately the Lagrange
multipliers $(\bfg{\lambda}^r)^T \equiv
(\lambda_1,\dots,\lambda_N)$ for the positions and
$(\bfg{\lambda}^p)^T \equiv (\lambda_{N+1},\dots,\lambda_{2N})$ for
the momenta. Then, Eq.\ (\ref{dynamicallagrange}) becomes
\begin{mathletters}
\label{hamiltonlagrange}
\begin{equation}
\bfg{\dot{\lambda}}^r - {\bf K} \cdot \bfg{\lambda}^p = 0,
\end{equation}
\begin{equation}
\bfg{\dot{\lambda}}^p + {\bf M}^{-1} \cdot \bfg{\lambda}^r = 0,
\end{equation}
\end{mathletters}
which can be combined to yield 
\begin{equation}
\label{dynamicallambdap}
{\bf M} \cdot \bfg{\ddot{\lambda}}^p = 
- {\bf K} \cdot \bfg{\lambda}^p. 
\end{equation}
We note that these dynamical equations for the momentum Lagrange multipliers
correspond to the equations of motion for a lattice of $N$ fictitious particles
with masses $m_n$ and instantaneous positions $\lambda_n^p$, interacting
via time-dependent harmonic forces. Combining Eqs.\ (\ref{finallagrange})
and (\ref{hamiltonlagrange}), we obtain the
corresponding boundary conditions
\begin{mathletters}
\label{dynamicallambdapbc}
\begin{equation}
\bfg{\lambda}^p(t_f) = \bfg{\Psi}^p \cdot [{\bf p}(t_f) - {\bf p}_f],
\end{equation}
\begin{equation}
\bfg{\dot{\lambda}}^p(t_f) = - {\bf M}^{-1}{\bf \Psi}^r \cdot [{\bf r}(t_f)
- {\bf r}_f],
\end{equation}
\end{mathletters}
where ${\bf \Psi}^p$ and ${\bf \Psi}^r$ are
$N$$\times$$N$-dimensional diagonal weight factor matrices with elements
$({\bf \Psi}^p)_{nl}=\delta_{nl}\psi_{n+N}$ and
$({\bf \Psi}^r)_{nl}=\delta_{nl}\psi_n$. Note, that these
boundary conditions for the Lagrange multipliers are specified at the
{\it final} time $t_f$. Thus their determination requires the knowledge
of the final positions ${\bf r}(t_f)$ and momenta ${\bf p}(t_f)$ of the
particles in the actual lattice.
We emphasize that the introduction of Lagrange multipliers only serves the
purpose of including the dynamics of the driven lattice as a constraint in
the objective functional (\ref{objfunc}). This results in coupled equations
for the dynamics of the actual lattice and that of the Lagrange multipliers.

The optimal external force is now obtained by minimizing the modified
objective functional (\ref{modobjfunc}) with respect to ${\cal F}(t)$.
This minimization can be done for an external force ${\cal F}(t)$ whose
time dependence is allowed to be arbitrary \cite{Shi88} or which has a
prescribed analytic form.\cite{Shi90} In view of the important aspect of
experimental feasibility, we discuss the latter approach. For ISRS
excitation, the laser frequency is neglected and we constrain ${\cal F}(t)$
to be a sequence of Gaussian pulses
\begin{equation}
{\cal F}^{\rm ISRS}(t) = \sum_i S_i e^{(t-t_i)^2/\Delta^2}
\end{equation}
with individual heights \{$S_i$\} and pulse center times \{$t_i$\},
and common width $\Delta$. For IR excitation, ${\cal F}(t)$ is taken to
have a linearly chirped IR frequency under a single Gaussian envelope
with fixed width:
\begin{equation}
\label{firpulse}
{\cal F}^{\rm IR}(t) = S e^{(t-t_0)^2/\Delta^2}
\sin(\theta + \omega t + \alpha t^2).
\end{equation}
In both cases, the external force depends on a set of variable parameters
\{$\tau_j$\}, namely $\{S_i\},\{t_i\}$ and $\Delta$ for the ISRS case,
and $S,t_0,\theta,\omega$, and $\alpha$ for the IR case.

The gradient of the modified objective functional with respect to the
external force parameters \{$\tau_j$\} has components
\begin{eqnarray}
\label{gradient}
&& {\partial \bar{J}[\bfg{\xi},{\cal F}(t,\{\tau_j\})] \over \partial
\tau_i} = \nonumber \\
&& \int_0^{t_f} dt\, \left[ \psi_{\cal F}{\cal F}(t,\{\tau_j\}) +
\left(\bfg{\lambda}^p\right)^T \cdot {\bf {\tilde q}}\right] 
{\partial {\cal F}(t,\{\tau_j\}) \over \partial \tau_i},
\end{eqnarray}
where Eq.\ (\ref{fdefinition}) has been used to write 
${\bfg \lambda}^T \cdot \partial {\bfg \phi}/\partial F(t) = 
({\bfg \lambda}^p)^T \cdot {\bf {\tilde q}}$.
The optimal control force ${\cal F}^{\rm opt}(t)$ is obtained by finding
the zero of this gradient. Following Ref.\ \onlinecite{Shi88}, we use an
iterative approach for this nontrivial numerical problem. Using an
educated guess for the force parameters \{$\tau_j$\}, we integrate
the dynamical equations (\ref{eomcompact}) for the actual particles'
driven motion forward in time from $t=0$ to $t=t_f$, starting from the
initial conditions (\ref{eombc}). This yields
${\bf K}(t)$ for $t$ in the interval $[0,t_f]$, as well as the final
positions ${\bf r}(t_f)$ and momenta ${\bf p}(t_f)$, which are used to
evaluate the boundary conditions (\ref{dynamicallambdapbc})
for the Lagrange multipliers at
$t=t_f$. We next integrate Eq.\ (\ref{dynamicallambdap}) backward
in time from $t=t_f$ to $t=0$. From this we obtain
$\bfg{\lambda}^p(t)$ for $t$ in $[0,t_f]$, which is then used
to evaluate the gradient components (\ref{gradient}). We adapted a
fifth-order Gear predictor-corrector molecular dynamics (MD) method
\cite{Allen87} for the time evolution of the particles in the actual
lattice and for the Lagrange multipliers. The force parameters are updated
within a conjugate gradient scheme,\cite{Buckley85} and the procedure is
repeated until the gradient (\ref{gradient}) is zero to within a specified
tolerance. 

For the case of purely harmonic potentials $V_{n,n-l}(r_n - r_{n-l})$,
$\bfg{\Phi}$ and ${\bf K}$ are time-independent matrices.
Consequently, for initial conditions ${\bf r}_i=0$ and ${\bf p}_i=0$
corresponding to the lattice at rest in its equilibrium configuration,
the exact optimal control force can be obtained by direct algebraic
manipulation of the matrix equations (\ref{dynamicaleq})
and (\ref{hamiltonlagrange}), without
iteration. This is detailed in Refs.\ \onlinecite{Shi90} and 
\onlinecite{Schwieters90} for the application to harmonic
molecules. Within this harmonic limit, we have used this algebraic
approach as an independent test of our iterative MD method described above.

The harmonic limit also allows considerations of the controllability of
the system. A system is said to be controllable if an arbitrary specified
target state $\bfg{\xi_f}$ can be reached exactly with a suitable
external force ${\cal F}(t)$. In Ref.\ \onlinecite{Schwieters90} it was 
pointed out that a harmonic system is completely controllable only if 
all normal
modes couple to the external force. Still, even if a system is not
controllable in this rigorous sense, i.e., the external force couples
only to a restricted set of normal modes, it may be possible to reach a
final state $\bfg{\xi}(t_f)$ that is close, although not exactly equal, to
the specified target state $\bfg{\xi}_f$. We will
return to this aspect in Sec.\ \ref{sec:opticalcreation} below.

\subsection{The model: interatomic potentials and characteristics}
\label{sec:basismodel}

We consider a 1D diatomic lattice with masses $m$ and $M$ ($>m$), where
nearest neighbors interact via Born-Mayer plus Coulomb (BMC) potentials
\begin{mathletters}
\begin{equation}
V_{mM}(r) = \lambda_{mM}e^{-r/\rho} - {q^2 \over r}
\end{equation}
\begin{equation}
V_{Mm}(r) = \lambda_{Mm}e^{-r/\rho} - {q^2 \over r},
\end{equation}
\end{mathletters}
while second neighbors interact via pure Coulomb potentials
\begin{equation}
V_{mm}(r) = V_{MM}(r) = {q^2 \over r}
\end{equation}
and more distant neighbors are assumed to be non-interacting.
Note that the interaction between an atom and its nearest neighbors
distinguishes between ``left'' and ``right'' neighbors. Accordingly,
minimization of the total potential energy of the static lattice leads
to asymmetric nearest-neighbor equilibrium separations $R_0^{mM}$ and
$R_0^{Mm} (\neq R_0^{mM})$. Although this asymmetry might appear unusual
at first sight, it indeed correctly represents the situation encountered
by mapping the collective motion of $(111)$ planes in a zinc blende
structure crystal onto an effective 1D model lattice.\cite{Roessler98} 
Furthermore, this asymmetry is in fact necessary to properly represent a 
lattice with first-order Raman active vibrational modes.

\begin{figure}
\epsfig{file=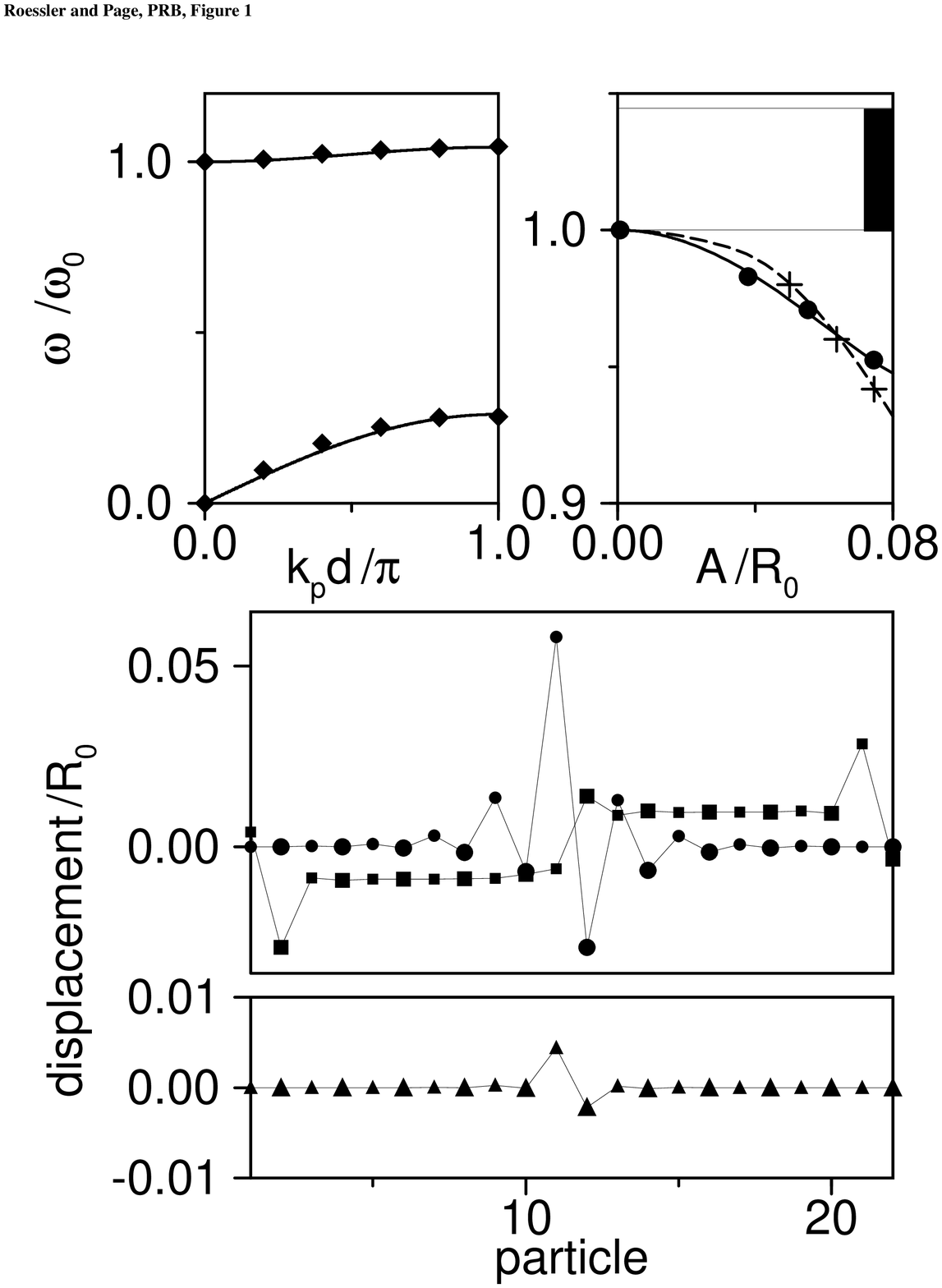,width=5.0in}
\caption{Characteristics of our 1D diatomic BMC model lattice
used for the optical creation of ILMs.
Upper left panel: harmonic dispersion (solid line) and experimental
transverse phonon frequencies (diamonds) along
$<$111$>$ in ZnS (Ref.\ \protect\onlinecite{Vagelatos74}).
Upper right panel: frequency vs amplitude curves for the optical
zone-center mode (OZCM) (solid line) and the related intrinsic localized
mode (OZCM-ILM) (dashed line) for 40 particles and standard periodic
boundary conditions (StdPBCs).
For the ILM curve the plotted amplitude is that of the mode's central
particle, which is a light mass. The thin horizontal lines locate the top
and bottom of the harmonic optical phonon band, indicated by the vertical
bar. The circles and crosses are the results of MD measurements of mode
frequencies for the OZCM and the OZCM-ILM, respectively. Our measurements
and RWA predictions differ by 0.5\% at most.
Middle panel: static (squares) and fundamental dynamic (circles)
displacements for an OZCM-ILM in a 22-particle diatomic BMC lattice with
free-end boundary conditions (FBCs). The small (large) symbols represent
the light
(heavy) masses. The lower panel gives the corresponding second harmonic
dynamic displacements as triangles.}
\label{model}
\end{figure}

In a detailed study,\cite{Bonart97} we noted that ILMs should be classified
according to their ``associated'' extended lattice mode (ExM), into which
they spatially broaden with decreasing amplitude.
Since we are focusing on the {\it optical} creation of ILMs, we study a
model that exhibits ILMs associated with the {\it optically-active} ExM. 
This intuitively reasonable choice will later turn out to be essential. 
With an external force such as given
by Eq.\ (\ref{externalforce}), the anharmonic version of the optical
zone-center mode (OZCM) with
displacement pattern $A(\dots,1,-m/M,1,-m/M,\dots)$ is both first-order
Raman and IR active, where for the former the asymmetry $R_0^{mM} \neq
R_0^{Mm}$ is necessary, as noted above. We therefore need to find model
parameters such that ILMs associated with the OZCM exist. In Ref.\
\onlinecite{Bonart97}, we obtained an ILM existence criterion based on the
interplay between fundamental harmonic and anharmonic dynamical properties
of the ILM's associated ExM. For our case of interactions via realistic BMC
potentials with their dominant soft anharmonicity, this criterion predicts
the existence of ILMs associated with the OZCM in diatomic lattices for
which the optical branch of the harmonic dispersion relation has a minimum
at $k=0$. In order to satisfy this constraint on the harmonic properties of
our 1D model lattice, while at the same time ensuring reasonably realistic
interatomic forces, we determined our BMC potential parameters by fitting
the harmonic dispersion to branches of the measured dispersion for a real
crystal along a high-symmetry direction. Since we consider both Raman and IR
excitation, we focused on real diatomic crystals with the zinc blende
structure, for which the ${\bf k}=0$ transverse optical (TO) phonon is both
first-order Raman and IR active. In particular, the measured TO phonon
branch of ZnS along the $<$111$>$ direction exhibits the required minimum at
${\bf k}=0$ (see Ref.\ \onlinecite{Vagelatos74}). Using ZnS masses 
$m=32.1\,$amu
and $M=65.4\,$amu, together with $\lambda_{mM}=3.45\times
10^2$eV, $\lambda_{Mm}=2.73\times 10^3$eV, $\rho=0.279$\AA, and $q=0.9e$,
we can approximately match those curves, as shown in the upper left panel
of Fig.\ \ref{model}. In the resulting model, $R_0^{mM}=1.67$\AA\ and
$R_0^{Mm}=2.95$ {\AA}, and the maximum of the harmonic phonon gap occurs at
the harmonic OZCM frequency $\omega_0=5.22\times10^{-2}$rad/fs.

The optimal control scheme described in Sec.\ \ref{sec:basisscheme} requires
the positions and momenta of a specified target state. In order to obtain
accurate predictions for the stationary solutions to Eq.\ (\ref{eom})
without external driving, we
use the well-established rotating wave approximation (RWA) for the
particles' time dependence,\cite{ILM} but generalized to include static
and second-harmonic contributions as well as oscillation at a mode's
fundamental frequency $\omega$. The motion of
particle $n$ is assumed to be of the form
\begin{equation}
\label{rwaansatz}
r_n(t) = b_n + c_n\cos(\omega t) + d_n\cos(2\omega t) + r_n^0.
\end{equation}
After inserting this ansatz into Eq.\ (\ref{eom}), we multiply the
resulting equations by either unity, $\cos(\omega t)$, or
$\cos(2\omega t)$, and average over a single period. For an $N$-particle
lattice, this yields a system of $3N$ coupled nonlinear equations for the
static displacements \{$b_n$\}, the fundamental dynamic displacements
\{$c_n$\}, and the second harmonic dynamic displacements \{$d_n$\}:
\begin {mathletters}
\label {averages}
\begin {equation}
0 = {1 \over 2\pi} \int_0^{2\pi}d\phi\, f_n(\{r_m\}),
\label {staticav}
\end {equation}
\begin {equation}
0 = m_n\omega ^2c_n +
{1 \over \pi}\int_0^{2\pi}d\phi\, \cos \phi\, f_n(\{r_m\}),
\label {fundamentalav}
\end {equation}
\begin {equation}
0 = m_n(2\omega) ^2d_n +
{1 \over \pi}\int_0^{2\pi}d\phi\, \cos (2\phi)\, f_n(\{r_m\}),
\label {secondharmav}
\end {equation}
\end {mathletters}
\noindent
where $\phi = \omega t$ and $f_n(\{r_m\})$ is defined in
(\ref{internalforce}). Once the boundary conditions are specified,
these equations can be solved using standard numerical routines; the
solutions in conjunction with Eq.\ (\ref{rwaansatz}) constitute the
RWA. We verify our RWA predictions by performing direct MD simulations
of Eq.\ (\ref{eom}), using a fifth-order Gear predictor-corrector
method.\cite{Allen87} Traditionally, Born-von Karman periodic boundary
conditions are employed for the description of bulk properties of
``infinite'' lattices. These boundary conditions are implemented by
setting $r_{n+N} \equiv r_n + L$, where $L$ is the static-lattice
equilibrium length of the supercell. Following the nomenclature of Ref.\
\onlinecite{Bonart97}, we denote them standard periodic boundary condition
(StdPBCs). While StdPBCs are convenient to describe infinite periodic
lattices, we will also consider finite systems with free-end boundary
conditions (FBCs).

Applying the
RWA to our model lattice, we find that ILMs associated with the OZCM exist,
as predicted by the criterion of Ref.\ \onlinecite{Bonart97}.
The ILM frequencies are in the
harmonic phonon gap. The middle panel of Fig.\ \ref{model} gives the
RWA static and fundamental dynamic displacements of such an OZCM-ILM at
$\omega=0.97\omega_0$ for a 22-particle system with FBCs. The
displacements are normalized to the average equilibrium
separation $R_0 \equiv (R_0^{mM}+R_0^{Mm})/2$. For clarity, the
corresponding second harmonic dynamic displacements, whose magnitude is
less than 8\% of the fundamental dynamic displacements, are shown
separately in the bottom panel. That this ILM is associated with a 
first-order Raman active ExM is reflected by the fact that its fundamental
dynamic displacement pattern exhibits no inversion symmetry. The static 
displacements seen at the ends of this finite system are the result of 
``surface'' relaxation. The upper right panel shows the RWA frequency 
vs amplitude curves
for the ILM and for the OZCM in the corresponding 40-particle lattice with
StdPBCs. For each of these two modes the amplitude is given by the 
magnitude of the largest fundamental dynamic displacement $c_n$. Also 
shown are the results of MD measurements of the ILM and OZCM frequencies,
which agree to within less than 0.5\% with the RWA predictions. The 
$\omega(A)$ curve for this ILM is indistinguishable from the corresponding 
curve for the 22-particle FBC lattice of the middle panel, as one would expect
from the mode's high localization.

\begin{figure}
\epsfig{file=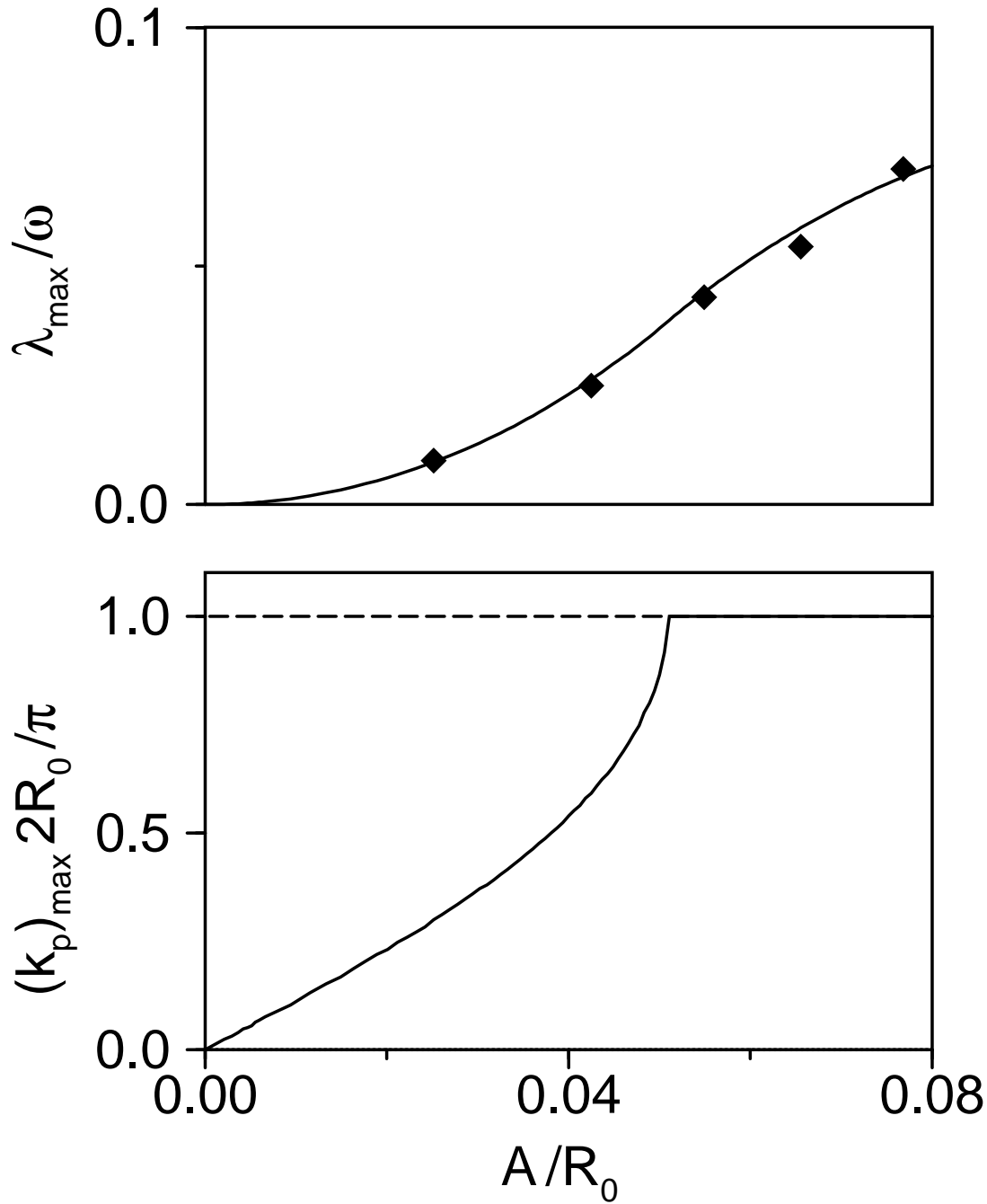,width=6.0in}
\caption
{Stability properties of the OZCM in our 1D diatomic BMC model lattice
with StdPBCs.
Upper panel: RWA-predicted maximum real instability growth rate of the
OZCM as a function of the normalized amplitude (solid line).
The diamonds give growth rate measurements obtained from MD simulations
for a 40-particle lattice. Lower panel: wave vector of the fastest-growing
Fourier component of the instability perturbation as a function of the
normalized amplitude. To achieve good resolution, the stability analysis
for both panels was based on a wave vector grid appropriate to a
2000-particle lattice.}
\label{modelstab}
\end{figure}

For our purpose of optical ILM excitation, the dynamical stability
properties of the optically-active ExM are important, as will be
seen later. We have therefore examined the stability of the OZCM in our model
lattice within an RWA-based approach detailed in 
Refs.\ \onlinecite{Sandusky94} and \onlinecite{Sandusky92}, but 
generalized to include perturbations of
the second harmonic contributions. This stability analysis assumes
infinitesimal displacement and velocity perturbations having an exponential
time dependence $\exp(\lambda t)$, leading to a linear eigenvalue
problem for the growth rates \{$\lambda$\}. In previous 
studies,\cite{Bonart97,Sandusky94} we have shown that a dynamical ExM
instability with a purely real growth rate $\lambda$ is intimately
connected with the existence of ILMs associated with the ExM. The OZCM in
our model lattice exhibits such an ILM-related instability. In the top
panel of Fig.\ \ref{modelstab}, we show the predicted maximum growth rate
as a function of the OZCM amplitude. MD measurements based on the
``projection method'' of Ref.\ \onlinecite{Sandusky94} agree 
to within 8\% with
the RWA predictions, as indicated by the diamonds. By decomposing the
instability perturbation into its spatial Fourier components, we can
extract the wave vector $(k_p)_{\rm max}$ of the fastest-growing component,
as discussed in Refs.\ \onlinecite{Bonart97} 
and \onlinecite{Sandusky94}. The bottom panel of
Fig.\ \ref{modelstab} plots $(k_p)_{\rm max}$ as a function of the OZCM
amplitude. At zero amplitude, $(k_p)_{\rm max}$ vanishes, and it increases
to its maximum allowed value $\pi/(2R_0)$ over a restricted range of
amplitudes. As discussed in Refs.\ \onlinecite{Bonart97} 
and \onlinecite{Sandusky94}, the
corresponding wavelength $2\pi/(k_p)_{\rm max}$ introduces a preferred
instability length scale at each amplitude. As the OZCM amplitude is
increased from zero, the instability length scale decreases from infinity,
reaching a maximum value of $4R_0$, after which it remains constant for
increasing amplitude. Finite-time MD simulations of unstable OZCMs seeded
with the fastest growing instability perturbation for
various amplitudes reveal that the instability leads to a breakup of the
OZCM into a periodic array of localized ILM-like excitations whose spacing
is very close to the preferred instability length. The stability properties
of the OZCM will be used in Sec.\ \ref{sec:indirectexcitation} below.

\section{Optical creation of ILMs}
\label{sec:opticalcreation}

Before we apply the optimal control scheme of Sec.\ \ref{sec:basisscheme} to
the optical creation of OZCM-ILMs in the model lattice described in Sec.\
\ref{sec:basismodel}, we consider the question of controllability, which
was briefly addressed in Sec.\ \ref{sec:basisscheme}. We find that this aspect
of our problem is different for different choices of the boundary
conditions.

For a {\it harmonic} diatomic lattice with StdPBCs, it is well
known that the harmonic version of the OZCM is the only normal mode that
couples to an optical-like force with a spatial dependence such as given
in Eq.\ (\ref{externalforce}). Hence, independent of the time-dependence of
${\cal F}(t)$, this external force can only excite the OZCM. On the other
hand, if we start from zero initial conditions in an {\it anharmonic}
lattice with StdPBCs, this argument still applies at short times, since
for the initially small amplitudes the interactions are dominated by the
harmonic terms. It remains true even at longer times for our StdPBC model
lattice, because the large-amplitude anharmonic OZCM continues to be an
exact stationary solution.

On the other hand, when a {\it finite} harmonic system with FBCs is
considered, the external force (\ref{externalforce})
couples with appreciable strength to a set of normal modes. However, the
size of this set decreases with increasing size of the system, and in the
limit $N\rightarrow\infty$ of an infinite system it is again just the FBC
normal mode corresponding to the OZCM in a StdPBC
lattice which can be excited optically. From these considerations of the
{\it harmonic} case, we expect the external force (\ref{externalforce}) to
have better control over a system with FBCs than over a lattice with
StdPBCs, when {\it
anharmonicity} is included. However, with FBCs we also expect to see a
dependence on the system size, with increased controllability for smaller
systems.

In Sec.\ \ref{sec:directexcitation} we show how ILMs can be excited
``directly'' in a finite system with FBCs. As expected, this approach
depends on the size of the system. For the creation of ILMs in crystal
lattices, we exploit the fact that the OZCM is unstable and breaks up into
ILM-like localized vibrations. Hence, although we can only excite the StdPBC
OZCM directly, its decay can produce ILM-like vibrations ``indirectly.''
This is detailed in Sec.\ \ref{sec:indirectexcitation}.

\subsection{``Direct'' excitation of ILMs in finite systems}
\label{sec:directexcitation}

\subsubsection{ISRS excitation}
\label{sec:directisrsexcitation}

\begin{figure}
\epsfig{file=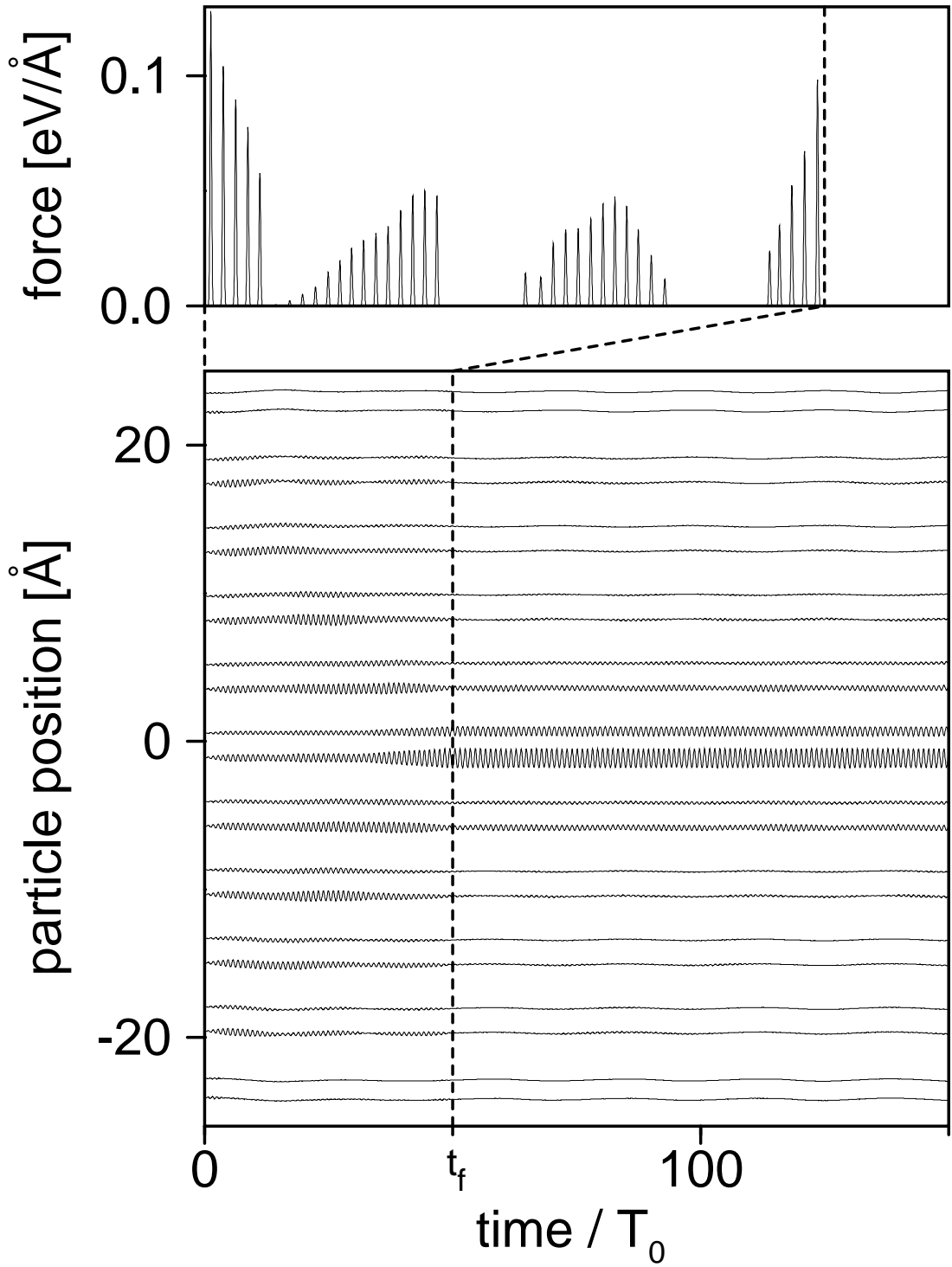,width=6.0in}
\caption
{Direct ILM excitation via ISRS.
Top panel: sequence of Gaussian pulses ${\cal F}(t)$ for the
direct creation of an ILM in a 22-particle system with free ends.
The bottom panel shows the MD results of applying this sequence, with the
particle displacements magnified by a factor 5. The applied field ends at
$t_f$. Note that the same force magnitude ${\cal F}(t)$ acts on each
particle.}
\label{ilmdirect}
\end{figure}

We first demonstrate the power of the optimal control scheme by showing
our results for ISRS excitation of an OZCM-ILM in a 22-particle system
with FBCs. Preliminary studies with simpler model systems revealed that
the control scheme becomes more successful for longer control periods and
larger numbers of pulses in the applied sequence. Keeping the
question of experimentally feasible laser pulse sequences in mind, we
choose a control interval $[0,t_f=50T_0]$, where $T_0 \equiv 2\pi/\omega_0$
is the period of the harmonic OZCM. At $t=0$ the particles are at rest at
their equilibrium positions. The system is then driven by a sequence of 49
Gaussian pulses. The MD time step used during the optimization for this
case and for all of the simulations in this paper is $T_0/100$. For the
target state at $t_f$
we specify the RWA-predicted positions and momenta of an ILM at frequency
$\omega = 0.97\omega_0$. The displacement pattern is given  
in the lower two panels of Fig.\ \ref{model}. Our initial studies with
simpler systems also showed that a combination of weight factors $\psi_n =
16(m_n\omega_0/2)$ and $\psi_{n+N} = 16/(m_n\omega_0/2)$ for the target
positions and momenta of particle $n$, respectively, and $\psi_{\cal F}=1$,
guarantees a good balance of the various terms in the objective functional.
Using these values in our control algorithm yields the pulse sequence
given in the top panel of Fig.\ \ref{ilmdirect}. The common full width at
half maximum (FWHM) $2\sqrt{\ln 2}\Delta$ of the pulses is 18 fs, and the
amplitudes \{$S_i$\} range from zero to 0.13 eV/\AA. The bottom panel shows
that the application of this rather complex sequence of pulses in MD
produces a strikingly ``simple'' result, namely the creation of a
highly-localized excitation which persists almost unchanged well after the
applied field ends at $t=t_f$. 

We note that the pulse sequence shown in the top panel of Fig.\ \ref{ilmindirect}
does not represent a unique solution of the minimization of the modified
objective functional (\ref{modobjfunc}). It turns out that depending on the
initial guess for the parameters of the sequence, the optimal control
algorithm reaches different local minima of (\ref{modobjfunc}) with distinct
values. We have not studied this aspect in detail, but among the
small sample of three different solutions we have obtained, the value of the
minimum corresponding to the sequence shown in the top panel of Fig.\
\ref{ilmdirect} was the smallest. The application of each of the three 
sequences in MD simulations produced a long-lived stationary localized 
excitation like that shown in Fig.\ \ref{ilmdirect}. From a practical 
viewpoint, the multiplicity of solutions suggests that it should be 
possible to impose additional constraints in the optimization algorithm to 
enhance the experimental feasibility of the resulting pulse sequence.

Although the results of Fig.\ \ref{ilmdirect} appear promising, they are
sensitive to the system's finite size as anticipated at the beginning of
this section. For instance, applying the identical pulse sequence to a
42-particle system with FBCs yields no localization. In Fig.\
\ref{ilmdirect}, the system's ends are essential for the flow of
vibrational energy towards the center to set up the target ILM. Hence 
this ``direct'' ILM excitation method is not suited for
crystals, although it may have relevance for the excitation of anharmonic
``local modes'' in molecules such as benzene (C$_6$H$_6$). Indeed, {\it ab
initio} MD simulations for benzene readily yield local modes,\cite{Lewis96}
and the addition of an optimal control scheme may allow the prediction of
optical waveforms for their creation.

\subsubsection{IR excitation}

We have attempted to obtain similar results for the direct IR excitation
of ILMs in finite systems, using a single chirped far-IR pulse of fixed
width for the external force. Although we can reach a final state
$\bfg{\xi}(t_f)$ that is quite well localized, the localization
does not persist for any significant time after the applied field ends.
Evidently, the more restricted prescribed analytic time-dependence of the
chirped Gaussian far-IR pulse reduces the ability for successful control.

\subsection{``Indirect'' excitation of ILMs in infinite lattices}
\label{sec:indirectexcitation}

The results of Sec.\ \ref{sec:directexcitation} indicate that due to its
size dependence, the direct excitation method is not a suitable approach for
the creation of ILMs in crystals. However, our previous studies of the
interrelation between ILMs and their associated ExM, as detailed in Refs.\
\onlinecite{Bonart97} and \onlinecite{Sandusky94}, suggest an 
alternative approach: because
we have designed our model such that gap ILMs are related to the
optically-active OZCM, we can create localized vibrations in the gap
``indirectly'' by optically driving the unstable OZCM.\cite{Schwarz99} For 
this, our choice of a model in which the ILMs' associated ExM is optically 
active is essential.

\subsubsection{ISRS excitation}
\label{sec:indirectisrsexcitation}

To illustrate, we again use ISRS excitation. For the optimal control
algorithm target state, we specify an OZCM of frequency $\omega=
0.98\omega_0$, in an 8-particle lattice with StdPBCs. This particular
system size is large enough to avoid computational complications due
to second-neighbor interactions across the supercell boundaries, and it is
sufficiently small to accelerate the optimal control scheme. The dynamics
of the anharmonic OZCM are independent of the size of the StdPBC supercell,
so that the resulting optimal fields apply to an infinite lattice. We
choose this particular target OZCM on the basis of our RWA stability
analysis of Sec.\ \ref{sec:basismodel}. At $\omega=0.98\omega_0$, the OZCM has
an amplitude $A=4.37\times10^{-2}R_0$, a maximum instability growth rate
$\lambda_{\rm max}=2.78\times10^{-2}\omega$, and the corresponding
preferred perturbation wave vector is $(k_p)_{\rm max}=0.624\pi/(2R_0)$.
Hence, this particular OZCM is expected to decay reasonably fast into
ILM-like localized excitations with an easily discernible spacing of about
$6.4R_0$. The advantage of choosing such an intermediate amplitude is
clear from Fig.\ \ref{modelstab}: at smaller amplitudes the
expected spacing is larger, but the corresponding growth rate is smaller,
and vice versa for larger amplitudes. The issue of the size of the growth
rate is important, because we want to ensure that the decay of the
unstable OZCM occurs on time scales where other processes, e.g., damping,
which are not included in our description of the lattice dynamics, will
not significantly alter the results. However, since an amplitude
corresponding to $\sim$4\% of the average equilibrium separation is quite
large, we have repeated the optimization using as a target state an OZCM of
smaller amplitude. The results for that case will be discussed at the end
of this section.

Starting from rest at $t=0$, the system is driven with a sequence of 49
Gaussian pulses over a control interval of $[0,50T_0]$, as for the direct
excitation detailed in Sec.\ \ref{sec:directisrsexcitation}. We again use
weight factors $\psi_n = 16(m_n\omega_0/2)$ and $\psi_{n+N} =
16/(m_n\omega_0/2)$, but the simpler control task here allows us to increase
the weight factor for the integrated square magnitude of the external force
to $\psi_{\cal F}=10$, without affecting the ability to reach the target
state. In contrast to the pulse
sequence given in Fig.\ \ref{ilmdirect} for the finite chain, the optimal
sequence for OZCM excitation is found to consist of pulses having nearly
equal amplitudes. Accordingly, we simplified the control algorithm so as
to vary the pulses' common width, common amplitude, and individual pulse
center times.

\begin{figure}
\epsfig{file=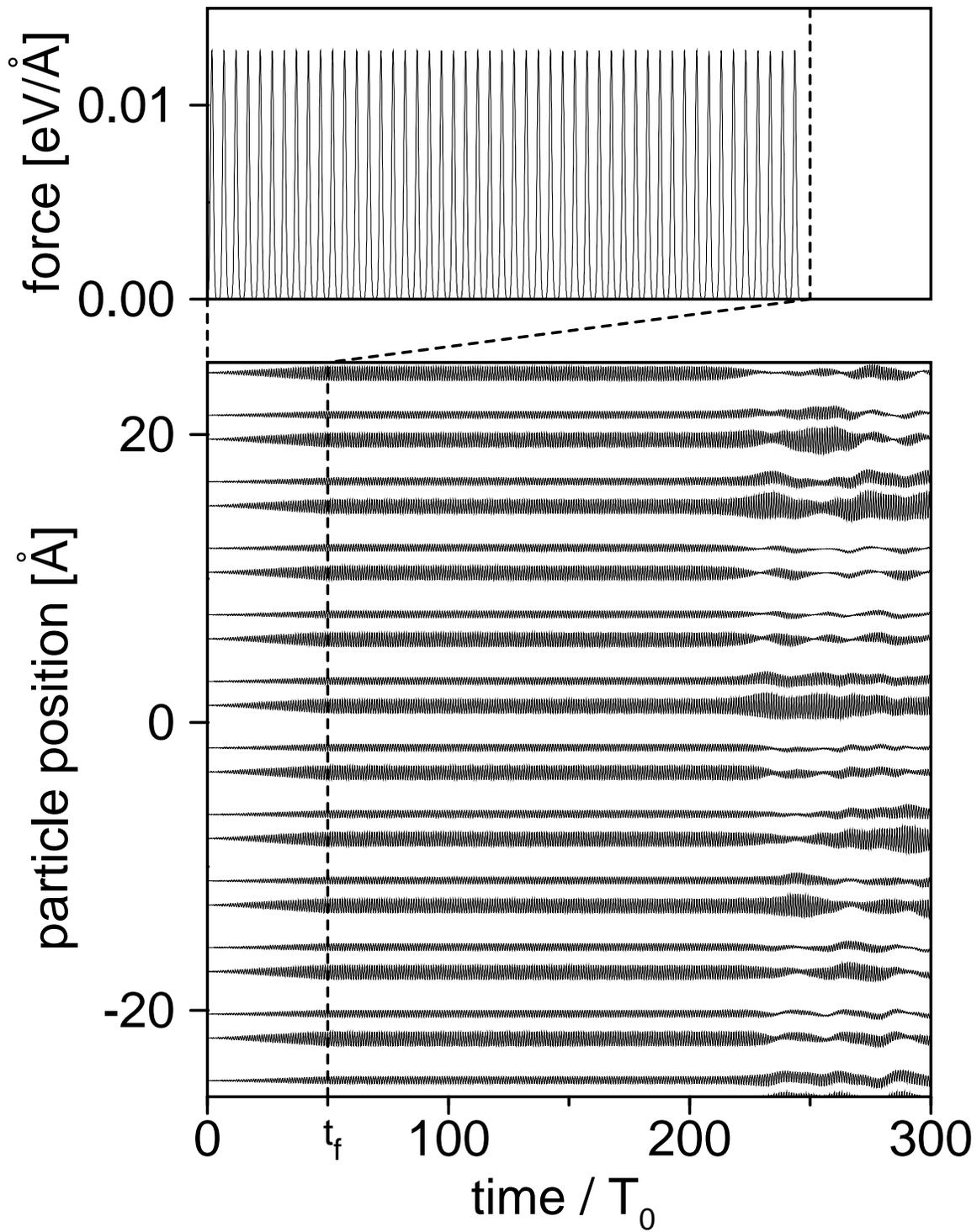,width=6.0in}
\caption
{Indirect ILM excitation via ISRS with T=0 initial conditions.
Top panel: sequence of Gaussian pulses ${\cal F}(t)$ for the
indirect creation of ILMs in a 40-particle lattice with periodic boundary
conditions. The bottom panel shows the resulting MD simulation, for zero
initial conditions. Displacements are magnified by a factor 5, and only a
portion of the lattice is shown.}
\label{ilmindirectt=0}
\end{figure}

The top panel of Fig.\ \ref{ilmindirectt=0} shows the
resulting ${\cal F}(t)$, which consists of pulses of FWHM 32 fs and
amplitude 0.013 eV/\AA. This is an order of magnitude less than the
largest amplitude for the direct ILM excitation in the finite 22-particle
lattice of Fig.\ \ref{ilmdirect}, and the equal amplitudes render this
sequence qualitatively simpler. However, a closer look reveals important
details, demonstrating how the control algorithm has globally
optimized this pulse sequence. The solid line in the upper panel of Fig.\
\ref{ozcmpulses} plots the position of a light particle in the OZCM as a
function of time during the last fifth of the control interval. The thin
vertical lines indicate the Gaussian pulse center times \{$t_i$\} of the
external force. As expected for an efficient impulsive driving force, the
pulse centers \{$t_i$\} coincide with the zero crossings of the particle
position. In addition, we measured the ``instantaneous'' amplitude of the
OZCM during the excitation by taking half the displacement difference
between adjacent turning points of the motion of a light particle and
assigning it the time of the intermediate zero crossing. Then we calculated
the corresponding undriven RWA frequency. This is shown as a solid line in
the lower panel of Fig.\ \ref{ozcmpulses}, while the circles indicate the
frequencies $2\pi/(t_{i+1}-t_i)$ corresponding to the spacings between
adjacent pulses of the external force. The good agreement between these
two quantities brings out the important (and somewhat hidden) aspect of the
sequence of Fig.\ \ref{ilmindirectt=0}: the spacing between adjacent
pulses varies through the sequence in such a way as to maintain resonant
impulsive driving of the anharmonic OZCM, whose frequency decreases as
its amplitude increases over the control interval. In Appendix A we show
that the qualitative aspects of this behavior can be readily understood by
doing the optimization for a sequence of $\delta$-function pulses.

\begin{figure}
\epsfig{file=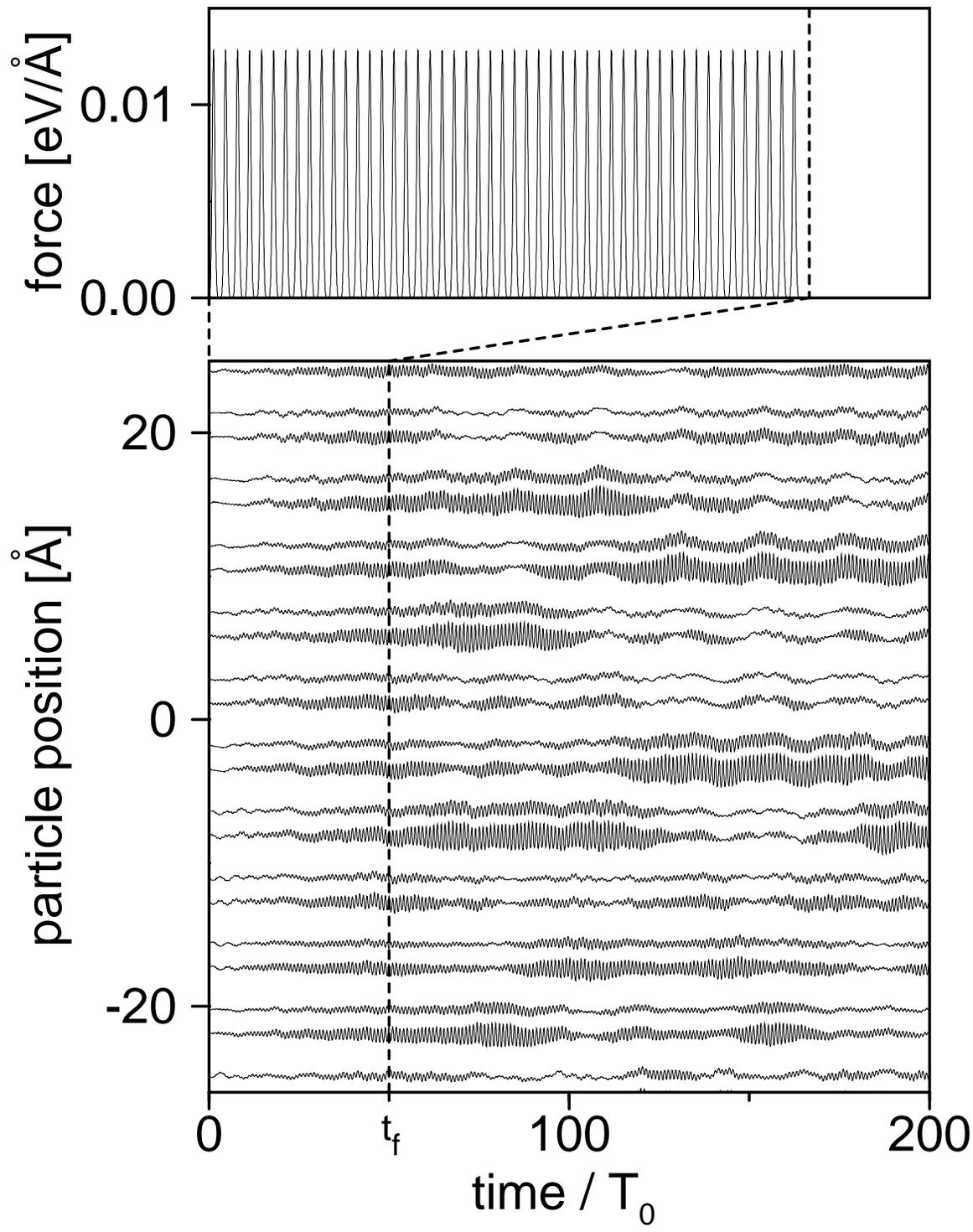,width=6.0in}
\caption
{Indirect ILM excitation via ISRS with T$>$0 initial conditions.
Top panel: same as top panel of Fig.\ \protect\ref{ilmindirectt=0}.
The bottom panel shows the resulting MD simulation, for random initial
velocities appropriate to a lattice temperature of 5 K. Displacements are
magnified by a factor 5, and only a portion of the lattice is shown.}
\label{ilmindirect}
\end{figure}

Applying our Gaussian pulse sequence to a 40-particle StdPBC lattice in an
MD simulation with the system initially at rest, we find that after the
field ends at $t_f=50T_0$, the excited OZCM keeps vibrating with constant
amplitude for about $150T_0$ until the perturbation due to accumulated
computational round-off error triggers the instability and the OZCM decays
into several localized excitations, as shown in the bottom panel of
Fig.\ \ref{ilmindirectt=0}. The spatial array of localized excitations is
not perfectly periodic because of the presence of instability perturbations 
of many wavevectors, with differing growth rates. Instead of relying on the 
clearly computer-dependent behavior of Fig.\ \ref{ilmindirectt=0}, we can 
provide the perturbation necessary
to trigger the OZCM instability by including the effects of nonzero
temperature. The bottom panel of Fig.\ \ref{ilmindirect} displays an MD
simulation for the same pulse sequence, but with random initial velocities
corresponding to a lattice temperature of 5 K. The presence of this
perturbation triggers the OZCM decay much sooner. Of course the details of
the MD results depend on the specific set of initial velocities, but for
ten sets consistent with 5 K we find the same qualitative results as shown
in Fig.\ \ref{ilmindirect}: the ILM-like localized excitations resulting
from the decay of the OZCM persist at fixed locations for several tens of
vibrational periods and tend to move slowly through the lattice. 

As mentioned earlier, we have repeated the indirect ILM excitation via ISRS
with the same control period, the same number of pulses in the sequence, and
identical weight factors, but using as a target state a less anharmonic OZCM 
at $\omega=0.99\omega_0$, with the corresponding amplitude $A=3.03 \times
10^{-2}R_0$. In this case, our optimal control algorithm yields a sequence
of pulses with 32 fs FWHM and amplitude 0.009 eV/\AA, compared with 32 fs
and 0.013 eV/\AA\ for the target OZCM at $\omega=0.98\omega_0$ and $A=4.37
\times 10^{-2}R_0$. The qualitative
time evolution of the unstable OZCM excited using this optimal sequence in
MD simulations with nonzero initial temperature is similar to that shown in
Fig.\ \ref{ilmindirect}, with the resulting ILM-like localized vibrations
having smaller amplitudes. Moreover, as expected from the larger preferred
instability length scale and smaller growth rate at this OZCM amplitude
(see Fig.\ \ref{modelstab}), the localized excitations are further apart
from each other and it takes roughly 20$T_0$ longer than in Fig.\
\ref{ilmindirect} before a comparable degree of localization is reached.

\subsubsection{IR excitation}
\label{sec:indirectirexcitation}

\begin{figure}
\epsfig{file=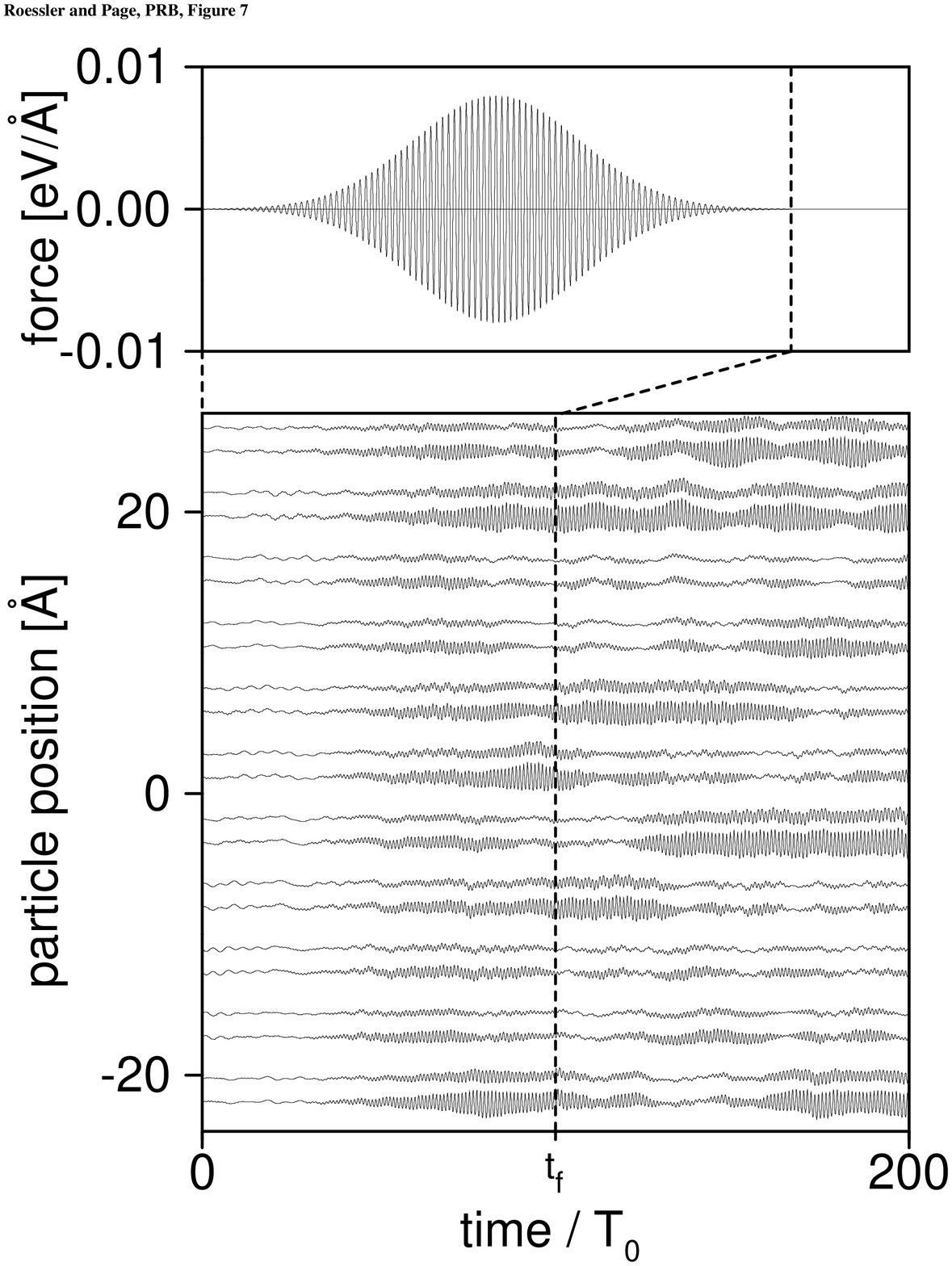,width=6.0in}
\caption
{Same as Fig.\ \protect\ref{ilmindirect}, but for the indirect ILM
excitation by a single, linearly chirped far-IR pulse.}
\label{ilmindirectir}
\end{figure}

We have also studied the indirect ILM creation via OZCM excitation using
IR, assuming for ${\cal F}(t)$ a single Gaussian pulse 
[Eq.\ (\ref{firpulse})] of fixed width
$33T_0$ (FWHM) and having a linearly chirped far-IR frequency over the
control interval $[0,100T_0]$. Using the same OZCM target state and weight
factors as detailed for the ISRS excitation above, the control scheme
yields an optimal force with pulse amplitude 0.008 eV/{\AA} and chirp rate
$-2.0\times10^{-7}$ fs$^{-2}$. The corresponding ${\cal F}(t)$ is shown
in the upper panel of Fig.\ \ref{ilmindirectir}, while the lower panel
displays the results of applying this external force to a 40-particle
StdPBC lattice in an MD simulation with initial random velocities
corresponding to a lattice temperature of 5 K. Just as for the ISRS
excitation of Fig.\ \ref{ilmindirect}, the external force excites a
slightly perturbed OZCM, which subsequently decays into ILM-like localized
excitations. The above discussion concerning different sets of random
velocities at the same temperature applies here as well.

Again, the optimization was repeated with the same control period, the same
fixed pulse width, and identical weight factors, but using as a target state
the smaller-amplitude OZCM at $\omega=0.99\omega_0$. In this case, our
algorithm yields an optimal force with pulse amplitude 0.006 eV/{\AA} and
chirp rate $-4.2\times10^{-8}$ fs$^{-2}$. In analogy to the excitation of
this OZCM with smaller amplitude via ISRS, the behavior in MD simulations
with nonzero initial temperature is qualitatively similar to that for the
larger-amplitude OZCM shown in Fig.\ \ref{ilmindirectir}, but exhibits the
same differences as in the case of ISRS excitation: the resulting localized 
vibrations are spatially further apart and it takes about 20$T_0$ longer 
until a comparable degree of localization is reached.

\section{Discussion}
\label{sec:discussion}

\subsection{Feasibility of the necessary external fields}

Having theoretically demonstrated the creation of ILMs using ``designer''
external forces, we now discuss the experimental feasibility of the 
corresponding fields. First we consider the excitation via ISRS. Pulse
shaping for ultrashort (13 fs) visible laser pulses has been 
demonstrated,\cite{Efimov95} with complicated final waveforms 
ranging from 12-pulse
sequences with an overall Gaussian envelope and equal spacing to 6-pulse
sequences with equal amplitudes and variable spacing. Furthermore, visible
lasers producing ultrashort (18 fs) pulses with extremely large field
magnitudes up to $\sim$270 V/\AA\ are available.\cite{Perry94,Barty96}
It is thus possible to produce the maximum field strengths of 1.22 V/\AA\
and 0.38 V/\AA\ necessary for the examples of Figs.\ \ref{ilmdirect} and
\ref{ilmindirect}, respectively, assuming a conservative value ${\cal P}=
2.5$ \AA$^2$ for the polarizability derivative.\cite{Calleja82} However,
the crucial experimental question is whether a given sample can
tolerate such high electric fields in an experiment.

In a {\it theoretical} argument,\cite{Yan85} Nelson and coworkers
estimated the potential of single-pulse ISRS to excite large-amplitude
anharmonic vibrations. Assuming pulses with 10 $\mu$J energy focused to 50
$\mu$m (FWHM) spot sizes, they predicted that a maximum phonon displacement
of $2\times10^{-3}$ \AA\ could be produced in the organic molecular crystal
$\alpha$-perylene. For pulse widths of 70 fs, these parameters correspond
to a field strength of about 0.5 V/\AA. Hence, their estimation of the
displacement produced by a {\it single} pulse and the field strength they
considered reasonable are comparable with the corresponding quantities in
our indirect ILM excitation via ISRS. Furthermore, they suggested that in
some materials coherent vibrational displacements in the 0.1-1 \AA\ range
could possibly be achieved. These optimistic predictions were preceded by
an actual experimental demonstration that single visible laser pulses of 70
fs width and 1 $\mu$J energy focused to 150 $\mu$m spot sizes impulsively
excited coherent optic modes in $\alpha$-perylene.\cite{deSilvestri85} The
field strengths, $\sim 0.05 V/\AA$, used in this single pulse experiment 
were one order of magnitude below those assumed for the subsequent 
theoretical prediction.  However, as discussed in a later experimental paper
by the same group,\cite{Weiner91} it turned out not to be feasible to use
the proposed larger field strengths, because they exceed the fairly low
laser-induced breakdown threshold of $\alpha$-perylene. This sequence of
publications highlights how a material's laser-induced breakdown threshold
can limit the potential of ISRS for the excitation of large-amplitude
vibrations. However, it was pointed out in Ref.\ \onlinecite{Weiner91}
that materials having a substantially larger laser-induced damage 
threshold than $\alpha$-perylene exist.

The question of breakdown thresholds in ultrashort pulse laser-solid
interaction is not yet very well studied. The breakdown thresholds for
alkali halides under irradiation by near-visible laser pulses with pulse
widths down to 10 ps were experimentally determined to be about 
0.2 V/\AA\ (Ref.\ \onlinecite{Bloembergen74}), with a tendency for the 
thresholds to increase with
increasing frequency and decreasing pulse width. Only few measurements for
femtosecond pulses, such as the pulses used in our ISRS excitation studies,
are available in the literature. For fused silica and the alkali fluorides,
breakdown thresholds above 0.5 V/\AA\ were obtained with 275 fs and 400 fs
pulses for visible and near-visible frequencies.\cite{Stuart96} Values
between 0.6 V/\AA\ and 1.0 V/\AA\ were measured for fused silica, sapphire,
magnesium fluoride, and glass, using visible laser pulses having a width of
120 fs.\cite{vdLinde96} Other experimental studies of fused silica
obtained breakdown threshold fields of 3.8 V/\AA, 3.0 V/\AA, and 3.3 V/\AA\
for visible pulses of 150 fs, 100 fs, and 55 fs width, 
respectively.\cite{Du94,Du96} We did not find measurements of 
the breakdown threshold in ZnS in the relevant short-pulse regime. However, 
an experiment on optical coatings made from ZnS (Ref.\ \onlinecite{Gu89}) 
showed that its breakdown
field strength for nanosecond pulses is comparable to that of magnesium
fluoride, for which the short-pulse values are given above. From these
experimental results for pulses that are still 2-20 times wider than the
ones we used for the excitation of ILMs via ISRS in Secs.\
\ref{sec:directexcitation} and \ref{sec:indirectexcitation}, it appears
that the maximum field strength for our direct ILM excitation of Fig.\
\ref{ilmdirect} for the 22-particle chain may be slightly too large to be
realized in an experiment, while that for the indirect ILM excitation in
the crystal lattice of Fig.\ \ref{ilmindirect} is below the values for
breakdown. As discussed at the end of Sec.\
\ref{sec:indirectisrsexcitation}, the necessary external force
magnitudes can be reduced by targeting an OZCM at a smaller amplitude,
but this approach has limitations due to the competition between the time
scales for the unstable OZCM decay and other processes in a crystal, e.g., 
damping. Moreover, the use of sequences with more pulses can also
decrease the force magnitudes: according to the $\delta$-pulse approximation
(see Appendix A), the force magnitude is inversely proportional to the
number of pulses in the sequence. However, since a larger number of pulses
requires longer control periods, the above {\it caveat} about the
competition of time scales applies here as well.

Turning to IR excitation, we note that the maximum force amplitude in
Fig.\ \ref{ilmindirectir} corresponds to a field strength 0.008 V/\AA,
assuming that $q=1.0e$. Free-electron far-IR lasers produce picosecond
pulses with intensities reported up to $4 \times 10^7$ W/cm$^2$ 
(Ref.\ \onlinecite{FEL}), corresponding to a field magnitude of 
0.002 V/\AA. Moreover, the frequency of free-electron far-IR lasers
can be chirped at rates of $-9 \times 10^{-9}$ fs$^{-2}$ 
(Ref.\ \onlinecite{Knippels95}). These field magnitudes and chirp 
rates are within a factor of 5 and 20,
respectively, of those used for the IR excitation of ILMs in Fig.\
\ref{ilmindirectir}. For this excitation mechanism, laser-induced
breakdown should not play a role as a limiting factor, since the threshold
of 0.2 V/\AA\ obtained for alkali halides using near-IR pulses of 10 ps
width \cite{Bloembergen74} is well above the necessary maximum field
magnitudes obtained here. As discussed at the end of Sec.\
\ref{sec:indirectirexcitation}, both the necessary field magnitudes and the
chirp rates are reduced when a smaller amplitude OZCM is targeted.

We conclude that the fields
necessary for the indirect excitation of ILMs via ISRS or far-IR as
demonstrated in Sec.\ \ref{sec:indirectexcitation} are reasonable and may be
feasible in the near future. Among the approaches we have considered,
excitation via ISRS seems more promising since lasers producing the
necessary high field strengths are already available, although the
experimental problem of laser-induced breakdown must be borne in mind.

Preliminary results
indicate that the indirect excitation of ILMs can also be achieved when
simpler analytic time dependences for the external force are used, without
necessitating significantly larger force magnitudes. For the indirect ILM
excitation via ISRS, we repeated the optimization procedure using pulse
sequences constrained to have an {\it equal} variable spacing between the
pulses as well as common variable widths and magnitudes and found that
for the target OZCMs at $\omega=0.98\omega_0$ and $\omega=0.99\omega_0$
the required force magnitudes were 2.3\% and 0.4\% larger, respectively.
Similarly, repeating the optimization for the indirect ILM excitation via
IR using a single Gaussian pulse with {\it unchirped} variable frequency as
well as fixed width and variable magnitude, we found the corresponding
increases in the force magnitude to be 5.0\% and 1.3\%.

Another important experimental consideration is the robustness of the
optimal fields. Considering the experimental limitations on the fidelity
of shaped waveforms for ISRS, we note from Ref.\ \onlinecite{Efimov95}
that pre-specified pulse amplitudes and widths were reproduced to within 
10\% and pulse positions to within 10 fs. Randomly perturbing the 
ISRS pulse parameters of the sequence shown in the top panel of Fig.\
\ref{ilmindirectt=0} for the indirect ILM excitation in the lattice within
these margins reveals that although the perturbed ${\cal F}(t)$ excites the
OZCM to a slightly different amplitude in each of the ten cases considered,
a decay into localized excitations always occurs. On the other hand, the
ISRS pulse sequence shown in the top panel of Fig.\ \ref{ilmdirect} for the
direct ILM excitation in the 22-particle chain is more sensitive to such
infidelities. In each of ten cases of randomly perturbing the ISRS pulse
parameters within the above margins, a localized excitation at the final
time results, but only in one case does this excitation persist after the
applied field is turned off. Decreasing the margins to a 5\% error for pulse
amplitudes and widths, and 5 fs for the pulse positions, the success rate
increases to four out of ten.

\subsection{Effects of nonzero initial temperature and damping}

We have obtained our optimal external forces assuming that the system is
initially at rest, but in an experiment thermal fluctuations will always be
present and we should consider their effect. For the case of indirect ILM
excitation, we have seen that the efficacy of the external field is enhanced
by the presence of velocity perturbations due to an initial temperature of 
5 K, since they serve to trigger the OZCM instability. At this low temperature,
thermal fluctuations are small enough to act just as perturbations on the 
zero temperature dynamics. Accordingly, application of the zero-temperature 
optimal force initially excites a slightly perturbed OZCM which subsequently
decays, as seen in the lower panels of Figs.\ \ref{ilmindirect}
and \ref{ilmindirectir}, respectively.

If we increase the initial temperature, the situation changes.
We have performed MD simulations with the ISRS pulse sequence of Fig.\
\ref{ilmindirectt=0} for ten different sets of initial velocities
appropriate to lattice temperatures of 77 K and 300 K. Already at 77 K
the initial excitation can no longer be identified as a perturbed OZCM.
Nevertheless, in all ten of the 77 K cases, the simulations result in 
well-localized ILM-like excitations, which appear sooner than for the 
5 K case of Fig.\ \ref{ilmindirect}.  However, fewer localized 
excitations are observed for the same size lattice, and an increased 
background of thermally excited long-wavelength acoustic vibrations 
is present. This trend continues as the initial temperature is raised 
to 300 K. Very similar results are obtained when the optimal force 
for the far-IR excitation from Fig.\ \ref{ilmindirectir} is used at 
higher initial temperatures. 
For comparison, we repeated the same nonzero initial temperature
simulations, but with no external force, and we observed no
significant localization of vibrational energy. Therefore, although
the optimal external force obtained for zero initial temperature
does not achieve its original goal of exciting an unstable OZCM when
used at these elevated temperatures, it nevertheless produces 
localized vibrations.


Similar to the question of robustness with respect to infidelities in the
pulse parameters, the optimal ISRS pulse sequence for the direct ILM
excitation in the 22-particle chain is more sensitive to nonzero initial
temperatures than is the sequence for indirect ILM excitation in the
lattice. Performing MD simulations with the ISRS pulse sequence of 
Fig.\ \ref{ilmdirect} for ten different sets of initial velocities 
appropriate to lattice temperatures of 5 K, we find that a persisting
localized excitation at the target site results in seven cases.
Already at an initial temperature of 77 K the success rate drops to zero,
although in some cases a localized excitation is created at a site other
than the target site.

Vibrations in real crystals couple to other types of excitations and
exhibit finite lifetimes -- typically between 5 and 300 phonon periods
for optical phonons.\cite{vdLinde93} We have included this aspect by
adding phenomenological velocity-dependent damping to our MD simulations. 
Anharmonicity, which contributes significantly to phonon lifetimes, is 
already treated explicitly in our calculations; thus we assume a small
damping constant corresponding to an OZCM lifetime of $100T_0$. We 
repeated the optimization for the direct ILM excitation via
ISRS of Fig.\ \ref{ilmdirect} after adding this damping, using the optimal
pulse sequence with zero damping as initial guess. This results in a
qualitatively very similar sequence of pulses, but with a larger maximum
force magnitude of 0.24 eV/{\AA}, compared with the earlier result 0.13
eV/\AA\ for zero damping. Due to the presence of damping the amplitude of
the resulting localized excitation decays away within a few tens of $T_0$
after the applied force ends. Similarly, when we include damping
in the optimization of the indirect ILM excitation via ISRS of Fig.\
\ref{ilmindirect}, we find that the common pulse amplitude increases from
0.013 eV/\AA\ to 0.016 eV/{\AA}. With $T=0K$ initial conditions, the
amplitude of the resulting OZCM damps out before the accumulated
computational round-off error can trigger the OZCM instability, but for an
initial temperature of 5 K, we again observe a break-up of the OZCM into
ILM-like localized excitations, whose amplitude subsequently damps out over
few tens of OZCM periods. Additional calculations for the indirect ILM
excitation via ISRS using a damping constant appropriate to a shorter
OZCM lifetime of $50T_0$ yield a common pulse magnitude 0.019 eV/\AA\, 
along with qualitatively similar MD results. Thus we conclude that it 
is possible to create ILMs with damping present, although the necessary 
force amplitudes are of course somewhat larger. Furthermore, detection 
of these ILMs would have to occur within few tens of OZCM periods after
their creation.

\subsection{Relevance for real crystals}

While our 1D model incorporates some realistic features, such as
standard interparticle potentials and the measured harmonic dispersion
of ZnS, it is not a model of any real crystal. We have also considered 
3D models of ZnS-structure crystals using standard two-body central 
potentials between atoms out to second neighbors.\cite{Roessler98}
It is well known that the harmonic modes for ${\bf k}$ along $<$111$>$
map onto an effective 1D model involving the collective motion of (111)
planes. We have shown that this mapping also occurs in the anharmonic 
case, for certain polarization directions. The resulting quasi-1D model, 
with effective anharmonic potentials between (111) planes undergoing 
collective motion with one transverse and one longitudinal degree of 
freedom, yields a representation of the actual motion in a 3D crystal 
and thus justifies our studies of purely 1D lattices.\cite{transinst} 
Applying again the ILM existence criterion of Ref.\ \onlinecite{Bonart97},
we can choose the parameters of the quasi-1D model such that
the ILMs are associated with the optically-active ExM. Hence, we adjusted
our potential parameters so as to fit the measured harmonic phonon
dispersion data of ZnS along $<$111$>$, as well as measured mode Gr\"uneisen
parameters. Within the RWA this model exhibits gap ILMs associated with 
the anharmonic version of the ${\bf k}=0$ TO phonon.\cite{Roessler98}
These results suggests that to the extent that the central potentials used
in this model capture the anharmonic properties of the real crystals,
representative candidate materials for the indirect optical excitation of 
these ILMs would be ZnS, ZnSe, and the copper halides. We re-emphasize that 
these candidates for indirect optical creation of ILMs have two basic
properties in common: (i) the ${\bf k}=0$ TO phonon in these materials is
both first-order Raman and IR active, and (ii) the frequency of the
${\bf k}=0$ TO phonon is at the {\it minimum} of the TO branch along
$<$111$>$, such that in conjunction with the fact that real potentials
are dominated by soft anharmonicity, the criterion of 
Ref.\ \onlinecite{Bonart97} predicts the existence of gap ILMs 
associated with this ${\bf k}=0$ TO phonon.

\section{Conclusion}
\label{sec:conclusion}

In conclusion, our optical excitation studies demonstrate theoretically
that suitably tailored laser radiation offers a promising route for the
laboratory creation of vibrational ILMs. The time dependence of the fields 
is determined by an efficient optimal control algorithm, designed to produce
waveforms consistent with the rapidly developing experimental capabilities
in laser pulse shaping. The {\it direct} excitation of ILMs in finite 
systems was demonstrated for impulsive stimulated Raman scattering 
by a sequence of ultrashort laser pulses at THz repetition rates, 
with variable spacing between consecutive pulses. For periodic lattices, 
ILM creation was achieved {\it indirectly} via decay of the unstable 
associated ExM which is excited optically either via multiple-pulse ISRS 
as above or via a single far-IR pulse with a linearly chirped frequency. 
For the indirect excitation approach, it is essential to consider a 
lattice having ILMs that are associated with the optically-active ExM.

The direct ILM excitation via ISRS requires rather complex pulse sequences
and laser field strengths that may just exceed the breakdown threshold of a
given sample. Furthermore, this approach shows a dependence on the system
size and is therefore not well suited for the creation of ILMs in crystal
lattices, although it may be relevant for the excitation of local
modes in molecules. Our studies show that the more advantageous means
to excite ILMs in crystals is via the decay of their associated unstable
anharmonic ExM, optically driven to a large amplitude. This approach not
only requires smaller field strengths, but it succeeds even in the presence
of thermal fluctuations and damping, and also with external forces
constrained to have simpler analytic time dependences, i.e., pulse
sequences with {\it equal} spacing between consecutive pulses for ISRS
and a single pulse with {\it constant} frequency for IR excitation. Hence,
our studies point to a potentially fruitful avenue for experimentally
accessing the regime of large-amplitude anharmonic vibrational dynamics,
which is very different than that for harmonic or weakly anharmonic systems.

\acknowledgements{Supported by NSF Grant No.\ DMR-9510182. J. B. Page 
also gratefully acknowledges the Max Planck Institute for the Physics of 
Complex Systems, Dresden, Germany, for their hospitality and support 
during the completion of the manuscript.}

\appendix
\section{Delta-approximation for ISRS excitation of the OZCM}
\label{app:delta}

In this Appendix we demonstrate that the main characteristics of the
optimal pulse sequence for the ISRS excitation of the OZCM in our model
lattice described in Sec.\ \ref{sec:indirectisrsexcitation} can be
reproduced using a simplification which was considered in 
Ref.\ \onlinecite{Cahn93}, but in a different context. Here, the 
external force due
to a sequence of short laser pulses is approximated by a sequence of
$L$ delta functions
\begin{equation}
\label{deltaforce}
{\cal F}(t) = \sum_{j=1}^L a_j\delta(t-t_j)
\end{equation}
with individual positive amplitudes $a_j$ and pulse center times $t_j$.
In addition, we know that the dynamics of the OZCM in a diatomic lattice
map onto that of an effective anharmonic oscillator with mass $m_{\rm eff}$,
displacement $x$ and momentum $p$. Including the external force
(\ref{deltaforce}), the Hamiltonian for the driven case is given by
\begin{equation}
\label{heff}
H_{\rm eff} = {p^2 \over 2m_{\rm eff}} + V_{\rm eff}(x) - x{\cal F}(t),
\end{equation}
where $V_{\rm eff}(x)$ is an effective anharmonic potential.

Starting with the effective oscillator initially at rest we want to excite
it to a given final energy $E_f$ while keeping the necessary external force
magnitude minimal. This requires an optimization of the parameters \{$a_j$\}
and \{$t_j$\} for the external force (\ref{deltaforce}). A single delta pulse
$a_j\delta(t-t_j)$ boosts the momentum of the oscillator by $a_j$ at time
$t_j$ and instantaneously changes the kinetic energy. Denoting the momentum
of the oscillator immediately before and after the pulse by $p_{j-1}$ and
$p_j$, respectively, we can write the energy boost as
\begin{equation}
\Delta E_j = {\left(p^2_j-p^2_{j-1}\right) \over 2m_{\rm eff}}=
{\left(a_j^2 + 2a_jp_{j-1}\right) \over 2m_{\rm eff}}.
\end{equation}
Evidently, for a given $a_j$ the largest possible energy transfer occurs
when the pulse arrives exactly at the time when $p_{j-1}$ is positive and
maximal, i.e., at the zero crossings of the displacement of the oscillator
where its momentum is positive. Knowing the effective potential, this
simple result is sufficient to determine the optimal pulse center times for
a given set of pulse amplitudes \{$a_j$\}. Furthermore, we now know that
the energy of the oscillator after the complete pulse sequence is given by
\begin{equation}
E_L = {1 \over 2m_{\rm eff}}\left(\sum_{j=1}^L a_j\right)^2.
\end{equation}
We define an objective functional
\begin{equation}
J(\{a_j\}) = \sum_{j=1}^L a_j^2 - \lambda\left(E_L - E_f\right),
\end{equation}
where the first term is just the integrated square magnitude of the
external force (\ref{deltaforce}) and the second term constrains the
energy $E_L$ after the pulse sequence to be equal to the desired final
energy $E_f$ via introduction of a Lagrange parameter $\lambda$.
Minimizing this objective functional with respect to the pulse amplitudes
\{$a_j$\}, we find that the optimal pulse sequence consists of pulses with
equal amplitudes
\begin{equation}
a_j = {1 \over L}\sqrt{2m_{\rm eff}E_f}.
\end{equation}

\begin{figure}
\epsfig{file=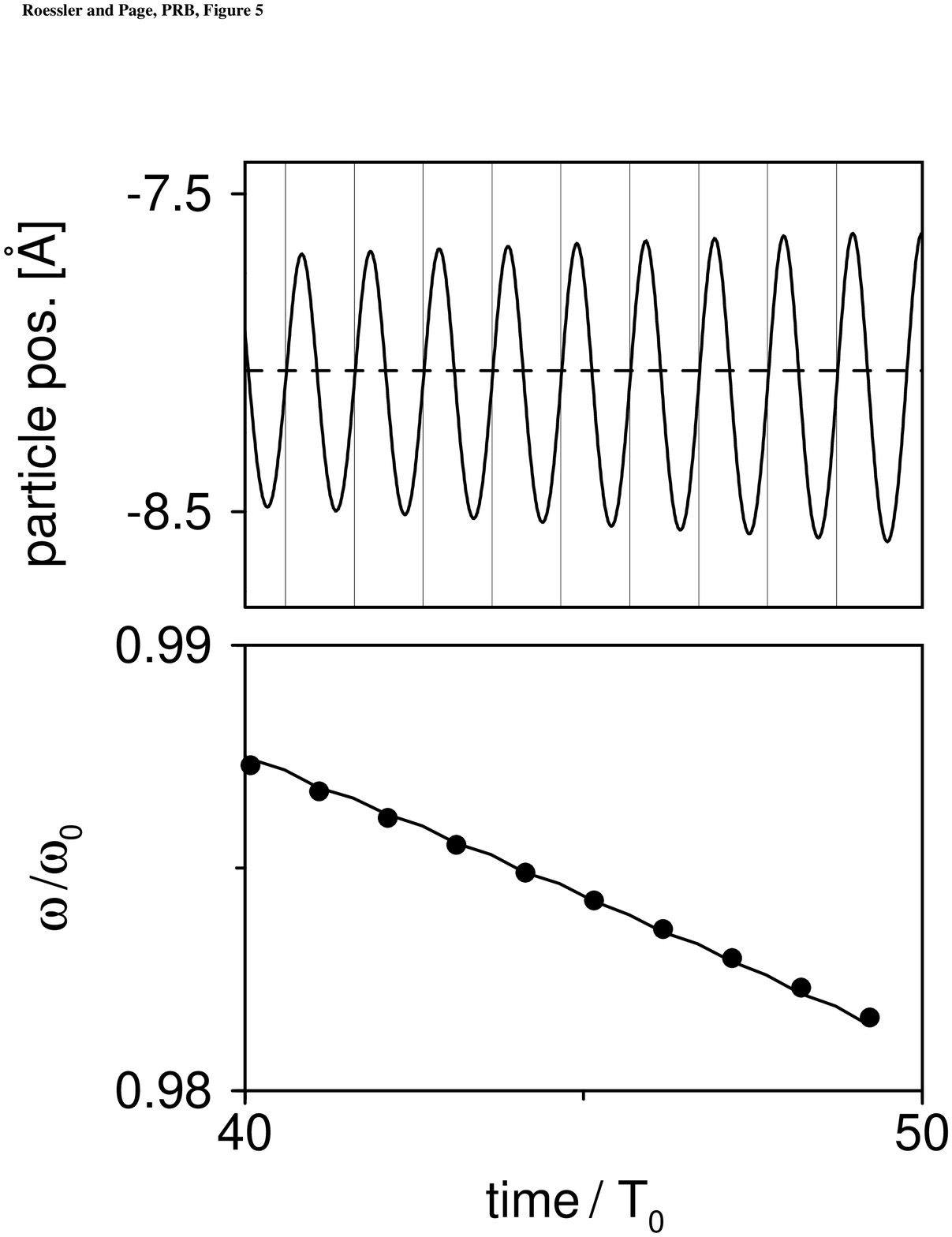,width=6.0in}
\caption
{Details of the ISRS pulse sequence for the indirect ILM excitation.
Top panel: position of a light particle during the OZCM excitation shown
in the bottom panel of Fig.\ \protect\ref{ilmindirectt=0} as a function of
time (solid line). Thin vertical lines denote the pulse center times $t_i$
of the corresponding optimal pulse sequence given in the top panel of Fig.\
\protect\ref{ilmindirectt=0}. Bottom panel: RWA frequency calculated from
the measured instantaneous OZCM amplitude during the excitation (solid
line). Circles indicate the frequencies $2\pi/(t_{i+1}-t_i)$.}
\label{ozcmpulses}
\end{figure}

To summarize, the $\delta$-approximation predicts that the optimal pulse
sequence for the OZCM excitation consists of pulses with equal amplitudes
whose pulse center times coincide with the zero crossings of the motion
of the atoms. This agrees with our numerical results from the full optimal
control scheme, as discussed in Sec.\ \ref{sec:indirectisrsexcitation} and
illustrated in Fig.\ \ref{ozcmpulses}. However, the $\delta$-approximation
underestimates the area under a single pulse by about 20\%, when compared
with the pulses in the numerically obtained optimal sequence. Hence, this
simplified description correctly predicts the qualitative features of the
optimal pulse sequence, but it cannot give quantitatively reliable results
for the optimal pulse width and magnitude. These require knowledge of the
actual motion of the particles over the duration of each pulse and are
obtained by applying the full optimal control scheme.

\clearpage

\onecolumn

\clearpage

\begin{references}
\bibitem{transinst} Our quasi-1D model does not address the possibility
of dynamical instabilities affecting the coherent motion of the 
lattice planes. If strong instabilities of this type
were to occur, they could considerably alter the dynamics in the 3D 
lattice from those predicted by the 1D model.
\end{references}
\end{document}